\documentclass[draft=false,BCOR=0mm,oneside,DIV=18]{scrartcl}

\usepackage[T1]{fontenc}
\usepackage[utf8]{inputenc}
\usepackage[english]{babel}
\usepackage{amsmath}
\usepackage{amssymb}
\usepackage{graphicx}   
\usepackage{siunitx}
\usepackage{comment} 
\usepackage{glossaries}

\newcommand{\diff}{\mathop{}\!\mathbin\mathrm{d}}
\newcommand*{\laplace}{\mathop{}\!\triangle}
\newcommand{\eref}[1]{eq.~(\ref{eq:#1})}
\newcommand{\fref}[1]{Fig.~\ref{fig:#1}}
\newcommand{\Figref}[1]{Figure~\ref{fig:#1}}
\newcommand{\mean}[1]{\langle #1 \rangle} 

\DeclareSIUnit[]
  \me{\mega\electronvolt}

\DeclareSIUnit[]
  \f{\femto\meter}

\DeclareSIUnit[]
  \nucl{nucl}

\newglossaryentry{adens_eln} { 
    name={\ensuremath{\mean{\glsentrytext{dens_eln}}}},
    description={average electron density}
}

\newglossaryentry{adens_byn} { 
    name={\ensuremath{\mean{\glsentrytext{dens_byn}}}},
    description={average baryon density}
}

\newglossaryentry{adens_ntn} { 
    name={\ensuremath{\mean{\rho_\textrm{n}}}},
    description={average baryon density}
}

\newglossaryentry{cpot_eln} { 
    name={\ensuremath{\mu_\textrm{e}}},
    description={electron chemical potential}
}

\newglossaryentry{cpot_eln_diff} { 
    name={\ensuremath{\Delta \glsentrytext{cpot_eln}}},
    description={difference of electron chemical potentials}
}

\newglossaryentry{cpot_eln_kin} {   
    name={\ensuremath{\glsentrytext{cpot_eln}^\textrm{kin}}},
    description={electron chemical potential}
}

\newglossaryentry{cpot_eln_un} { 
    name={\ensuremath{\glsentrytext{cpot_eln}^\mathrm{c}}},
    description={electron chemical potential of uniform electron distribution}
}

\newglossaryentry{cpot_ntn} { 
    name={\ensuremath{\mu_\textrm{n}}},
    description={neutron chemical potential}
}

\newglossaryentry{cpot_prn} { 
    name={\ensuremath{\mu_\textrm{p}}},
    description={proton chemical potential}
}

\newglossaryentry{dens_byn} { 
    name={\ensuremath{\rho_\mathrm{B}}},
    description={baryon density}
}

\newglossaryentry{dens_eln} { 
    name={\ensuremath{\rho_\textrm{e}}},
    description={electron density}
}

\newglossaryentry{dens_ntn} { 
    name={\ensuremath{\rho_\mathrm{n}}},
    description={neutron density}
}

\newglossaryentry{dens_nnucl} { 
    name={\ensuremath{\rho_0}},
    description={normal nuclear density}
}

\newglossaryentry{dens_prn} { 
    name={\ensuremath{\rho_\mathrm{p}}},
    description={proton density}
}

\newglossaryentry{dens_tc} { 
    name={\ensuremath{\rho_\mathrm{Q}}},
    description={total charge density}
}

\newglossaryentry{en_bind} { 
    name={\ensuremath{B(\glsentrytext{num_ntn},\glsentrytext{num_prn})}},
    text={\ensuremath{B}},
    description={binding energy}
}

\newglossaryentry{en_kin} { 
    name={\ensuremath{E_\mathrm{kin}^\mathrm{r}}},
    description={kinetic energy}
}

\newglossaryentry{en_kin_un} { 
    name={\ensuremath{E_\mathrm{kin}^\mathrm{c}}},
    description={kinetic energy of uniform electron distribution}
}

\newglossaryentry{en_sep_prn} { 
    name={\ensuremath{S_\mathrm{p}
        (\glsentrytext{num_ntn},\glsentrytext{num_prn})}},
    text={\ensuremath{S_\mathrm{p}}},
    description={separation energy of protons}
}

\newglossaryentry{en_nucl} { 
    name={\ensuremath{E_\mathrm{nucl}}},
    description={total nuclear energy}
}

\newglossaryentry{en_nucl_un} { 
    name={\ensuremath{\glsentrytext{en_nucl}^\prime}},
    description={total nuclear energy of uniform electron distribution}
}

\newglossaryentry{en_wsc} { 
    name={\ensuremath{E^{\ast}}},
    description={Wigner-Seitz cell energy}
}

\newglossaryentry{en_wsc_diff} { 
    name={\ensuremath{\Delta \glsentrytext{en_wsc}}},
    description={difference of Wigner-Seitz cell energy}
}

\newglossaryentry{end_kin} { 
    name={\ensuremath{\tau_\mathrm{B}}},
    description={kinetic energy density}
}

\newglossaryentry{eln_frac} { 
    name={\ensuremath{Y_\mathrm{e}}},
    description={electron-to-baryon fraction}
}

\newglossaryentry{f_mom} { 
    name={\ensuremath{k_\mathrm{F}}},
    description={Fermi momentum}
}

\newglossaryentry{ffrac_wsc} { 
    name={\ensuremath{X}},
    description={Wigner-Seitz cell filling fraction}
}

\newglossaryentry{rat_ptb} { 
    name={\ensuremath{Y_\mathrm{p}}},
    description={proton-to-baryon ratio}
}

\newglossaryentry{mass_eln} { 
    name={\ensuremath{m_\mathrm{e}}},
    description={electron mass}
}

\newglossaryentry{num_byn} { 
    name={\ensuremath{A}},
    description={number of baryons}
}

\newglossaryentry{num_ntn} { 
    name={\ensuremath{N}},
    description={number of neutrons}
}

\newglossaryentry{num_prn} { 
    name={\ensuremath{Z}},
    description={number of protons}
}

\newglossaryentry{pot_els} { 
    name={\ensuremath{U_\textrm{e}}},
    description={electrostatic potential}
}

\newglossaryentry{vol_wsc} { 
    name={\ensuremath{V_\mathrm{c}}},
    description={Wigner-Seitz cell volume}
}

\newglossaryentry{rad_wsc} { 
    name={\ensuremath{R_\mathrm{c}}}, 
    description={Wigner-Seitz cell radius}
}

\title{Nuclear structure calculations for neutron-star crusts}
\author{
Claudio Ebel\footnote{ebel@fias.uni-frankfurt.de} \, \textsuperscript{1)}
\and
Thomas~J.~Bürvenich \footnote{thomas.buervenich@tergau-walkenhorst.com} \,
\textsuperscript{1)}
\and
Igor Mishustin\footnote{mishustin@fias.uni-frankfurt.de} \,
\textsuperscript{1,2)}
}
\publishers{1) \textit{Frankfurt Institute for Advanced Studies, J.W. Goethe
University, Ruth-Moufang Strasse 1,
60438 Frankfurt am Main, Germany}\\
2) \textit{National Research Centre Kurchatov Institute, 123182 Moscow,
Russia}}
\date{\today}
\begin{document}
\maketitle
\begin{abstract}
The goal of this paper is to investigate properties of clusterized nuclear
matter which is believed to be present in crusts of neutron stars at subnuclear
densities. It is assumed that the whole system can be represented by the set of
Wigner-Seitz cells, each containing a nucleus and an electron background under
the condition of electroneutrality. The nuclear structure calculations are
performed within the relativistic mean-field model with the NL3
parametrization. The first set of calculations is performed assuming the
constant electron background. The evolution of neutron and proton density
distributions was systematically studied along isotopic chains until very
neutron-rich system beyond the neutron dripline. Then we have replaced the
uniform electron background with the realistic electron distributions, obtained
within the Thomas-Fermi approximation in a self-consistent way with the proton
distributions. Finally, we have investigated the evolution of the
$\beta$-stability valley as well as neutron and proton driplines with the
electron density.
\end{abstract}
\section{Introduction}
The properties of clusterized nuclear matter at subnuclear densities and
moderate temperatures are under active investigation for several decades,  for
a review see refs. \cite{Pethick1991, Glendenning2000, Haensel2007}. After the
seminal papers \cite{Baym1971, Negele1973}, this topic became a very active
research field in recent years. Many microscopic and macroscopic models have
been used, see e.g. refs \cite{Gogelein2007, Grill2011, Than2011, Sharma2015}.
The superfluid properties of such inhomogeneous matter have been studied in
refs \cite{Baldo2007, Pastore2011, Jin2017}. Often the calculations are done
within the Wigner-Seitz approximation, when the system is split in
electrically-neutral cells containing a nucleus and electrons.
\par
In our previous paper \cite{Burvenich2007}, we have studied the influence of a
uniform electron background on nuclear ground states and decay modes. The
nuclear structure calculations were performed within the Relativistic Mean
Field (RMF) model with parameterization NL3 \cite{Lalazissis1997}. It was found
that for large electron Fermi momenta, $\alpha$ decay and spontaneous fission
are significantly altered, and, depending on the electron density, lead to
stabilizing or destabilizing effects. With growing electron density, the
$\beta$-stability valley moves to more neutron-rich systems, and the proton
drip line moves to more proton-rich nuclei. These changes happen because the
single-particle potential for protons gets deeper as electrons are added. Since
neutrons are electrically neutral, in the first approximation the neutron drip
line is unaffected by the presence of electrons. It was found that for electron
Fermi momenta of $k_F \approx \SI{0.05}{\per \f}$, the $\beta$-stability line
reaches the neutron drip line, and free neutrons occur. Hence, for larger
electron densities, these quasi-free neutrons need to be taken into account.
\par
In the present paper, we extend this model in two directions. We include the
dripping-out neutrons which populate unbound states. As before, we adopt the
Wigner-Seitz approximation representing the whole system as a collection of
non-interacting electrically-neutral cells characterized by the nuclear charge
$\gls{num_prn}$, baryon number $A$, the $\gls{num_prn} / \gls{num_byn}$ ratio
and average baryon density $\gls{adens_byn} = \gls{num_byn} / \gls{vol_wsc}$,
where $\gls{vol_wsc} = ( 4 \pi \gls{rad_wsc}^3 ) / 3$ and $\gls{rad_wsc}$ is
the WS cell radius. We employ boundary conditions for the single-particle
states which are physically motivated and which provide a non-zero density of
the neutron gas at the cell boundary.
\par
In \cite{Ebel2015} we have shown that using realistic electron distributions
obtained by solving the Poisson equation leads to a significant reduction of
the electrostatic energy and electron chemical potential. In the present work,
we have calculated the electron density distributions consistent with the
proton distributions obtained self-consistently within the RMF model. We
calculate the whole nuclear chart up to very heavy nuclei which can be formed
in neutron star crusts. Both nuclear matter and the electron distributions in
a WS cell are calculated using an iterative algorithm. We demonstrate that the
realistic electrostatic potentials induce a shift of the $\beta$-stability
valley and the proton drip line to more proton-rich systems thus utilizing the
stabilizing effect of the electrons. The differences in the electron chemical
potentials are evaluated for different values of $\gls{num_prn}
/ \gls{num_byn}$ and the baryon density.
\section{The Framework}
\subsection{The RMF model}
The nuclear structure calculation below are carried out within the same RMF-NL3
model as in ref. \cite{Burvenich2007}, but now the unbound neutrons are allowed
too. The corresponding Lagrangian has the form
\begin{multline}
    \mathcal{L} = \sum_\alpha \omega_\alpha \bar{\psi}_\alpha ( i \gamma_\mu
    \partial^\mu - m_N ) \psi \\
    + \frac{1}{2} \partial_\nu \sigma \partial^\nu
    \sigma - \frac{1}{2} m_\sigma^2 \sigma^2 - \frac{b}{3} \sigma^3
    - \frac{c}{4} \sigma^4 - g_\sigma \sum_\alpha \omega_\alpha \sigma
    \bar{\psi}_\alpha \psi_\alpha \\
    - \frac{1}{4} \omega_{\mu \nu} \omega^{\mu
    \nu} - \frac{1}{2} m_\omega^2 \omega^\mu \omega_\mu - g_\omega \sum_\alpha
    \omega_\alpha \omega^\mu \bar{\psi}_\alpha \gamma_\mu \psi_\alpha \\
    - \frac{1}{4} \vec{\rho}_{\mu \nu} \cdot \vec{\rho}^{\mu \nu} - \frac{1}{2}
    m_\rho^2 \vec{\rho}^\mu \cdot \vec{\rho}_\mu - g_\rho \sum_\alpha
    \omega_\alpha \vec{\rho}^\mu \cdot \vec{\psi}_\alpha \gamma_\mu \vec{\tau}
    \psi_\alpha \\
    - \frac{1}{4} F_{\mu \nu} F^{\mu \nu} - e \sum_\alpha \omega_\alpha A^\mu
    \bar{\psi}_\alpha \gamma_\mu \frac{1 + \tau_3}{2}\psi_\alpha
    \text{\,,}
\end{multline}
where $\alpha$ denotes the nucleons (neutron or proton) and $\omega_\alpha$ are
the occupation probabilities of the nucleon states. They take into account the
BCS pairing with a density-indepentent $\delta$-force \cite{Bender2000}. This
approach allows us to treat both bound and unbound nucleons on the same level.
In this paper we assume zero temperature.
\subsection{The Wigner-Seitz approximation}
The calculations below are performed within the Wigner-Seitz (WS) approximation
by dividing the system into WS cells, each containing only one nucleus and the
number of electrons equal to the nuclear charge $\gls{num_prn}$. The
introduction of WS cells is aimed as a convenient approximation of an ensemble
of nuclei surrounded by a more or less uniform background of electrons and
neutrons. Its construction should ensure that the physics contained in an
individual cell is to large degree independent of the surrounding matter. This
requires, for example, charge neutrality of the cell. However, there could be
also other requirements which minimize the electrostatic interactions with
neighbouring cells (see below). The accuracy of the WS approximation for the
inner neutron-star crust has been discussed, e. g. in  Refs. \cite{Baldo2006,
Chamel2007}. We use the units with $\hbar = c = 1$.
\subsection{Boundary conditions for unbound neutrons}
\label{boundary}
In our calculations we allow the possibility that some neutrons are not bound
in a nucleus, but occupy the whole available volume and reach the boundary of
the WS cell, which in our scheme coincides with the edge of the numerical grid.
For this reason, boundary conditions for the neutron states at the cell edge
need to be specified. The choice of these boundary conditions should be guided
by physical considerations. Physically, the unbound neutrons are expected to
form a more or less homogeneous gas, in which the nuclei are embedded. Thus,
the neutron density at the edge of the cell should have a value which coincides
to a large degree with the average density of unbound neutrons. Futhermore, the
slope of the neutron density should be zero in order to have a continuous gas
in the region between the cells and to allow a continuous connection of
physical quantities across cell boundaries.
\par
In the literature, different possibilities have been discussed. As demonstrated
in \cite{Negele1973}, reasonable properties of the neutron states can be
achieved if the boundary conditions are chosen differently for states with odd
and even angular momenta. For spherical cells, such a a sorting of states is
possible since the angular momentum is a conserved quantity in a spherically
symmetric system. One possible choice is to set
\begin{equation}
    \psi_l \left(r = \gls{rad_wsc} \right) = 0
    \text{, $l$  odd};
    \quad
    \frac{\diff}{\diff r} \psi_l \left( r = 0 \right) = 0
    \text{, $l$ even.}
\end{equation}
This way, approximately half of the states assumes a zero value at the
boundary, while the other half assumes a zero slope. There is no physical
reason why the condition above should be chosen this way and not the other way
round, i.e., that states with odd angular momentum are required to have zero
slope and states with even angular momentum are required to have zero
value. Baldo et al \cite{Baldo2006} have investigated the effect of a different
choice on physical oberservables and found a non-neglegible effect.
\par
As has been shown already in \cite{Negele1973}, these boundary conditions are
often not sufficient to prevent a building up of density at the cell edge
during the numerical iterations. One possible reason for such a buildup during
the iterative procedure is given by the fact that -- as many neutron states are
involved -- many states have large angular momenta and are thus located at
rather large radii. By this reason, the density accumulates near the cell edge,
giving rise to an attractive mean field potential for other neutron states.
Therefore, during the iterations, more and more neutron states are attracted to
the border of the cell.
\par
In \cite{Negele1973}, this problem has occured in the framework of
non-relativistic mean-field calculations. It was cured by an averaging
procedure; i.e. both the particle density $\gls{dens_byn} (r)$ and the kinetic
energy density $\gls{end_kin} (r)$ were averaged over a radial region in the
cell where the neutron gas density was approximatey constant. The densities in
a region close to the grid egde with $r = 1/\gls{f_mom}$ were then substituted
with these values. In each iteration, the density close to the cell edge was
brought back to the value $\gls{adens_ntn}$ of the average neutron gas in the
cell
\begin{equation}
    \gls{adens_ntn} = \frac{3 \gls{num_ntn}}{4 \pi \gls{rad_wsc}^3}
    \text{\,.}
\end{equation}
\par
We checked this averaging procedure in our calculations and made several test
runs. While the procedure worked in principle, it lead to some undesired
effects. For some nuclei and cell radii, this procedure led to a well-behaved,
smooth and almost constant neutron gas within the cell. For certain cases,
however, the averaging near the cell edge cured the buildup of density in each
iteration but did not lead to a reduction of the density at the cell edge during
the iterative procedure. In each new iteration density was still summing up at
the cell edge. If such a condition remained until the convergence of the
solution, the net effect resulted in a removal of neutrons from the cell, since
the large rise of the neutron density at the cell edge was replaced by average
values, and no measure was taken to conserve neutron number. In extreme cases,
a reduction of the neutron number by up to $15 \%$ was found. These artifacts
showed up in observables such as the binding energy of the system. Since the
particle number in spherical calculations scales with $\propto r^2$, even small
changes of the density at the boundary can lead to a large error.
\par
To conserve the particle number and get rid of these side effects, a different
method was designed and implemented in our present calculations. The behavior
of the neutron gas near the cell edge has been constrained by the physical
condition that the neutron density should have a zero slope at the cell edge.
Such constraints in self-consistent calculations can be used in all kinds of
situations where some nuclear properties should have prescribed values. For
example, the fission barrier of a heavy or superheavy nucleus can be computed
by putting a constraint on the quadrupole moment of the nucleus and by varying
its value. In our case, the constraint was chosen such that the neutron gas
density was required to have zero slope at the boundary. It was implemented by
simply requiring that the difference of the values of the neutron density at
the two outer-most grid points should be zero, i.e.
\begin{equation}
    0 = \gls{dens_ntn} (r = \gls{rad_wsc})
    - \gls{dens_ntn} (r = \gls{rad_wsc} - \Delta_r),
\end{equation}
where $\Delta_r$ is the radial spacing of the numerical grid in coordinate
space. In principle, more constraints could be added regarding other pairs of
grid points near the border of the grid, but with respect to the desired speed
and convergence of the calculations this most simple version proved to be most
adequate.
\par
In fact, this constraint can be applied not only to the neutron density but to
the total baryon density as well. In most cases, the protons are located at the
cell center and form bound states. Then the proton wave functions are
approximately zero at the cell edge, and the proton density does not contribute
to the total baryon density there. In cases when the protons dissolve and the
proton states reach out to the cell edge, such a requirement is fully
justified. The use of this constraint in our calculations led to satisfactory
results in most cases. The particle number was always exactly conserved. Only
for very large neutron numbers and/or comparatively small cell radii or large
average values of the neutron density, a small density bump close to the cell
edge appeared. The constraint was employed during the whole iterative procedure
or switched on at a later stage, depending on the system size and the
convergence speed.
\subsection{Initialization of states}
\label{initialization}
For the considered extremely neutron-rich nuclear systems, the initialization
proved to be quite important. During a self-consistent iteration of a regular
bound nucleus, more states than nucleons in the system are usually initialized,
and the lowest states are occupied with nucleons. During the iterative
procedure, however, non-occupied states can become energetically more favorable
than the occupied states, in which case these states will be occupied with
nucleons. For systems with a moderate number of states, this scheme works
extremely well and provides the configuration with the lowest ground-state
energy.
\par
For systems with many neutrons, a straight-forward initialization of all states
from scratch can be problematic. Firstly, the states are usually initialized in
a harmonic oscillator well, which has only bound states. Thus, these states are
quite different from the unbound states which they will become during the
iterative procedure. Therefore, rapid changes of these states will occur during
the first few iterations. Secondly, for the ground-state configuration, many
states should be bound within the nuclear potential.
\par
For these reasons, for the calculation of a nuclear system with $\gls{num_prn}$
protons and $\gls{num_ntn}$ neutrons,  where $\gls{num_ntn} \gg \gls{num_prn}$,
the following iterative procedure was used: in a first step, a nucleus with
$\gls{num_prn}$ protons and $\gls{num_ntn}^{\prime}$ neutrons was calculated,
where $\gls{num_ntn}^\prime \geq Z$ corresponds to a neutron number for which
the nucleus is close to the line of stability. After the solution has
converged, the proton and neutron states (both occupied and unoccupied) are
stored. In a second step, these states are used for the initialization of
a nucleus with $\gls{num_prn}$ protons, but $\gls{num_ntn}^{\prime \prime}$
neutrons, where $\gls{num_ntn}^{\prime \prime} > \gls{num_ntn}^\prime$.
Relations $\gls{num_ntn}^{\prime \prime} = \gls{num_ntn}^\prime + 2$ or
$\gls{num_ntn}^{\prime \prime} = \gls{num_ntn}^\prime + 4$ proved to work well.
After such a chain of calculations, a well-behaved solution for the extremely
neutron-rich system with $\gls{num_prn}$ protons and $\gls{num_ntn} \gg
\gls{num_prn}$ neutrons could be obtained.
\subsection{Simplified treatment of electrons}
Following our previous work \cite{Burvenich2007} for rough estimates, we employ
in this section a simplified model assuming a uniform electron density.
Generally, the chemical potential of the electrons can be represented as
\begin{equation}
    \gls{cpot_eln} = \gls{cpot_eln_kin} + \gls{pot_els}
    \text{.}
\end{equation}
The kinetic part is simply given by
\begin{equation}
    \gls{cpot_eln_kin} = \sqrt{\gls{f_mom}^2 + \gls{mass_eln}^2} 
\end{equation}
where the electron Fermi momentum $\gls{f_mom}$ does not depend on $r$. The
second contribution to the chemical potential stems from the electrostatic
potential $\gls{pot_els}$ in which they move. This potential is produced by the
total charge density $\gls{dens_tc} (r)$, which is given by both electrons and
protons
\begin{equation} \label{eq:dens_tc}
    \gls{dens_tc} = e [\gls{dens_prn} (r) - \gls{dens_eln} (r)]
    \text{\,.}
\end{equation}
The electrostatic potential is thus space-dependent due to the spatial
dependence of the proton density. Thus, for given constant $\gls{cpot_eln}$ and
space-dependent $\gls{pot_els}$, $\gls{f_mom}$ would have to be space-dependent,
too, which is a contradiction to the assumption of a constant electron
background. This inconsistency can be cured by some extent by introducing
a constant potential $\gls{pot_els}^0$, defined as the average electric
potential felt by an electron.
\par
Since $\gls{pot_els}$ is defined as
\begin{equation}
    \gls{pot_els}(r_1) =
    \int \diff^3 r_2 \frac{e^2 \gls{dens_tc} (r_2)}{|r_1-r_2|}
    \text{\,,}
\end{equation}
the average electric potential can be calculated as
\begin{equation}
    \gls{pot_els}^0 = \frac
    {\int \diff^3 r \gls{pot_els}(r) \gls{dens_eln} (r)}
    {\int \diff^3 r^\prime \gls{dens_eln} (r^\prime)}
    \text{\,,}
\end{equation}
where $\int \diff^3 r^\prime \gls{dens_eln}(r^\prime) = \gls{num_prn}$ is the
total number of electrons (and due to charge neutrality) equal to the number of
protons in the system. Thus, in the calculations presented in this section we
use the expression
\begin{equation}
    \gls{cpot_eln} = \gls{cpot_eln_kin} + \gls{pot_els}^0
    \text{\,.}
\end{equation}
The condition of $\beta$-stability is given by
\begin{equation} \label{eq:beta_stab}
    \gls{cpot_prn} + \gls{cpot_eln} = \gls{cpot_ntn}
\end{equation}
For zero temperature, the Fermi energies of protons and neutrons correspond to
their respective chemical potentials. The Fermi energies are obtained from the
iterative calculation. When pairing is employed, the Fermi energy is not
uniquely determined but lies somewhere between two states such, that the
average nucleon number is fulfilled. Due to these uncertainties, we formulate
this condition as
\begin{equation}
    \left| \gls{cpot_prn} + \gls{cpot_eln} - \gls{cpot_ntn} \right| < \Delta
    \text{\,,}
\end{equation}
where $\Delta \approx \SI{1}{\me}$.
\section{Results}
\subsection{Evolution of density profiles}
\subsubsection{Isotopes}
\begin{figure*}[t]
    \centering
    \includegraphics[width=.8\textwidth]
        {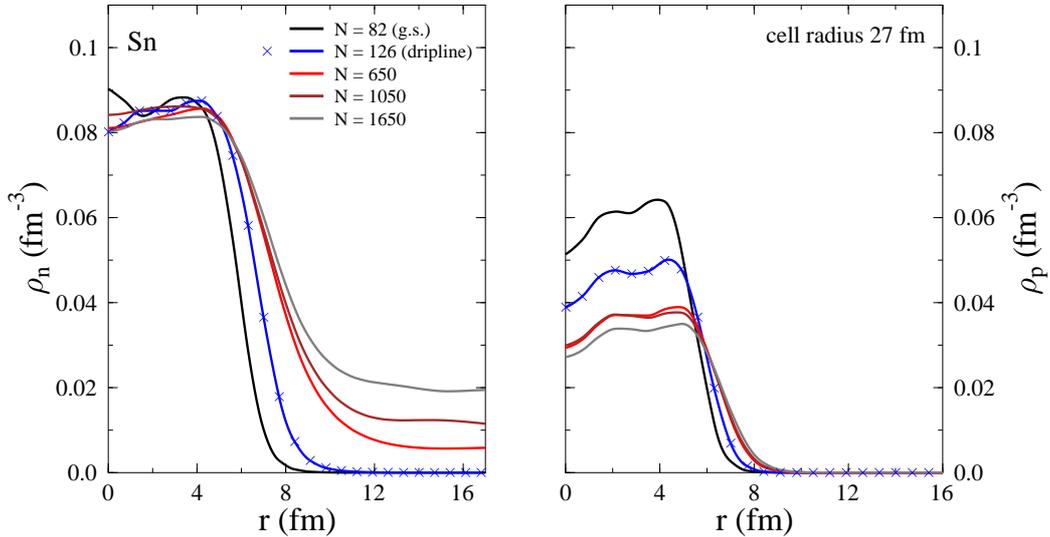}
    \caption{%
        Neutron (left) and proton (right) single-particle densities for
        a sequence of tin isotopes in a Wigner-Seitz cell with radius
        $\gls{rad_wsc} = \SI{27}{\f}$. The configuration corresponding to the
        neutron drip line is indicated by the crosses.}
    \label{fig:tin_isotopes}
\end{figure*}
\begin{figure*}[t]
    \centering
    \includegraphics[width=.8\textwidth]
        {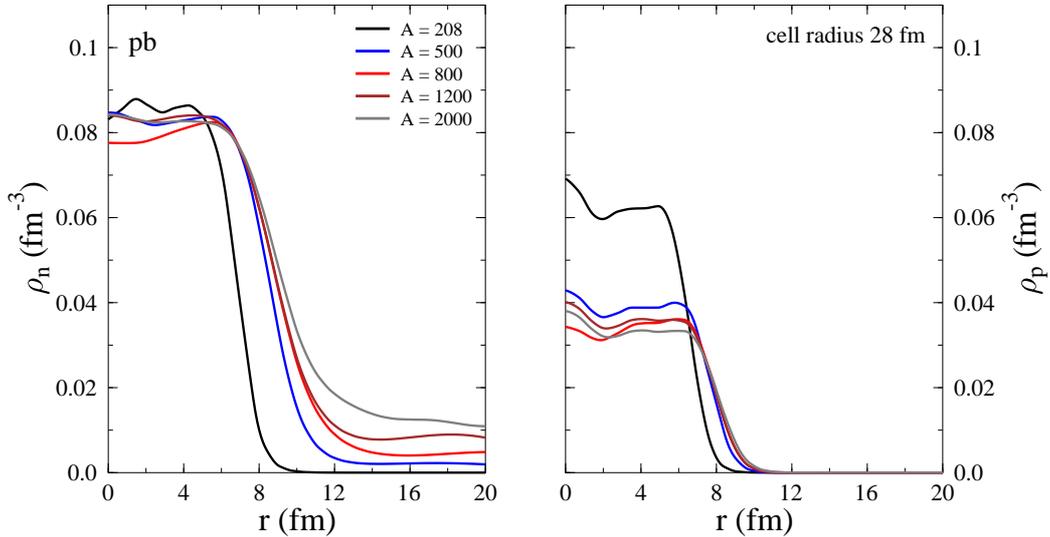}
    \caption{%
        Neutron (left) and proton (right) single-particle densities for
        a sequence of lead isotopes in a Wigner-Seitz cell with radius
        $\gls{rad_wsc} = \SI{28}{\f}$.}
    \label{lead_isotopes}
\end{figure*}
\begin{figure*}[t]
    \centering
    \includegraphics[width=.8\textwidth]
        {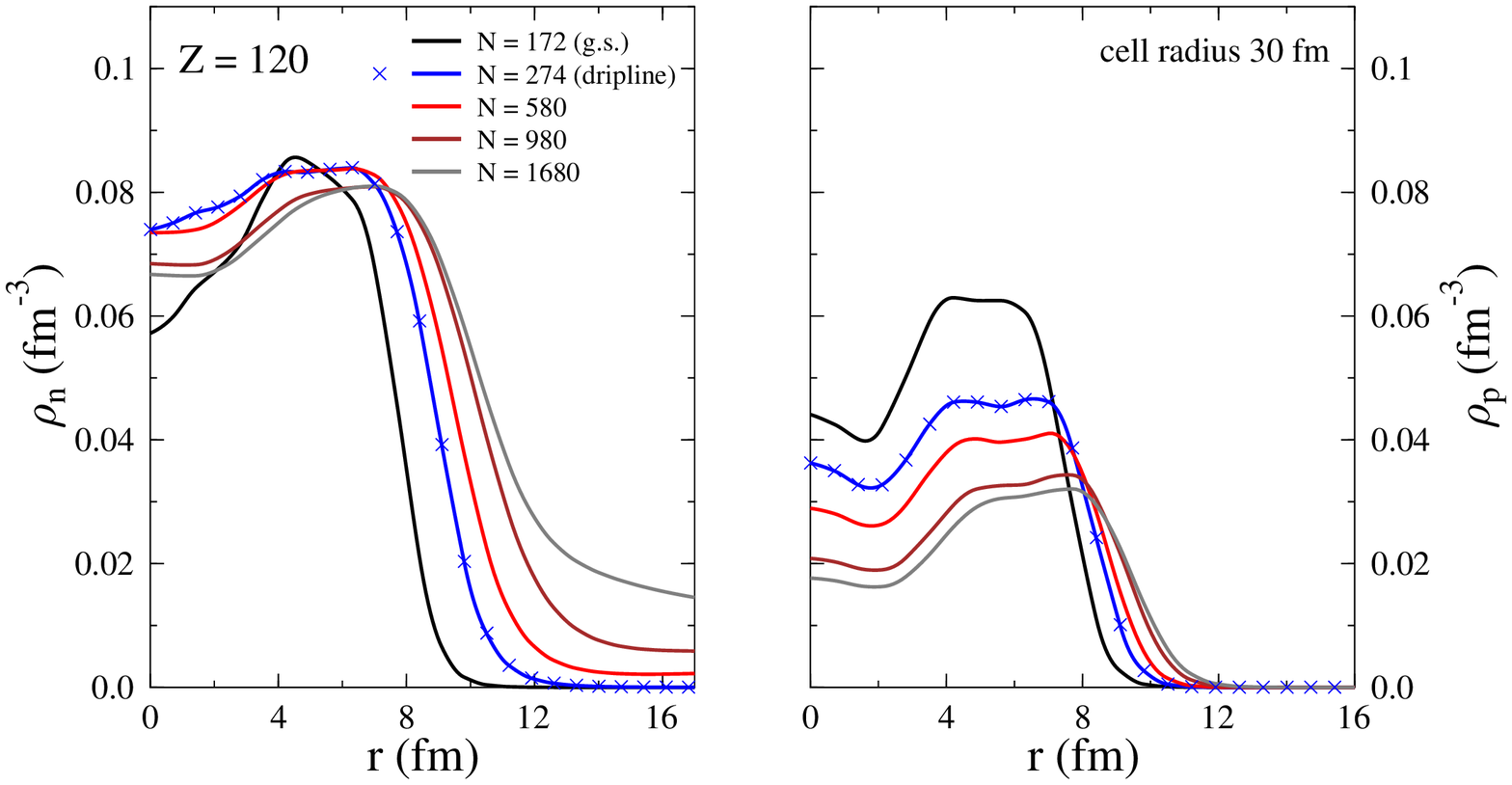}
    \caption{%
        Neutron (left) and proton (right) single-particle densities for
        a sequence of $\gls{dens_prn} = 120$ isotopes in a Wigner-Seitz cell
        with radius $\gls{rad_wsc} = \SI{30}{\f}$.}
    \label{120_isotopes}
\end{figure*}
First we investigate the structural evolution of shapes along isotopic chains.
It is assumed that the nucleus is located at the center of the spherical
Wigner-Seitz cell of radius $\gls{rad_wsc} = \SI{27}{\f}$. We start with
a $\beta$-stable isotope and add neutrons. In \fref{tin_isotopes} left and
right shows, respectively, the neutron and proton densities for tin isotopes.
The drip-line isotope is marked by crosses. These isotopes correspond to
different total baryon numbers within the cell and thus to different baryon
densities.
\par
Up to the drip-line isotope with $\gls{num_ntn} = 126$, the neutron density
shifts to larger radii, but all neutrons are bound. Beyond the drip line,
neutrons leak out from the nuclear potential and start to form a constant
neutron gas filling the cell approximately uniformly. For $\gls{num_ntn}
= 1650$, the neutron density at the cell edge is approximately $\gls{dens_ntn}
\approx \SI{.02}{\per \f \cubed}$. The corresponding average total baryon
density in the cell is $\gls{dens_byn} = \SI{.0206}{\per \f \cubed}$. These
numbers are practically equal because i) $\gls{num_ntn} \gg \gls{num_prn}$ and
ii) the dripped neutrons are distributed more or less uniformly in a big
volume. Also note that the central neutron density remains almost constant for
all isotopes, $\gls{dens_ntn} \approx \SI{.085}{\per \f \cubed}$, the value
corresponding to the saturated nuclear matter.
\par
For all neutron numbers, the protons remain localized within the central part
of the cell. With increasing neutron number, however, their distribution
broadens. In contrast to the neutron distributions, the central density of the
protons changes with increasing neutron number. While it is $\gls{dens_prn}
= \SI{.06}{\per \f \cubed}$ for the beta-stable isotope, the central proton
density drops below $\gls{dens_prn} = \SI{.04}{\per \f \cubed}$ for the
heaviest isotope.
\par
In our opinion, this behavior can be explained by the symmetry energy
contribution. Indeed, when the dripped neutrons build up a significant density
outside the nucleus, it is energetically favorable to stretch the proton
distribution in order to reduce the neutron-proton asymmetry in the transition
region. This stretching effect is rather robust. It is clearly seen in all
three considered cases. According to our knowledge, this effect was never
discussed previously. One should also bear in mind that the increased asymmetry
in the central region is less important in the energy balance because of the
$r^2$ weighting of different spherical shells.
\par
Analogous plots for lead isotopes ($\gls{num_prn} = 82$) and isotopes of the
superheavy element $\gls{num_prn} = 120$ are shown in Figs. \ref{lead_isotopes}
and \ref{120_isotopes}, respectively. The neutron distributions in both cases
are rather similar to the one for Sn isotopes. But the proton distributions
show large differencies. In the case of $\gls{num_prn} = 82$, the protons have
a small bump at the center. But in the case of the superheavy system
$\gls{num_prn} = 120$, protons have a pronounced central deep, which goes down
to $\SI{.02}{\per \f \cubed}$ for $\gls{num_ntn} = 1650$. According to our
knowledge, such hollow shapes have not been discussed previousely. In our
opinion, this is a consequence of the strong Coulomb repulsion of protons in
such a heavy system.
\subsubsection{Constant electron to baryon fraction}
\begin{figure*}[t]
    \centering
    \includegraphics[width=.8\textwidth]
    {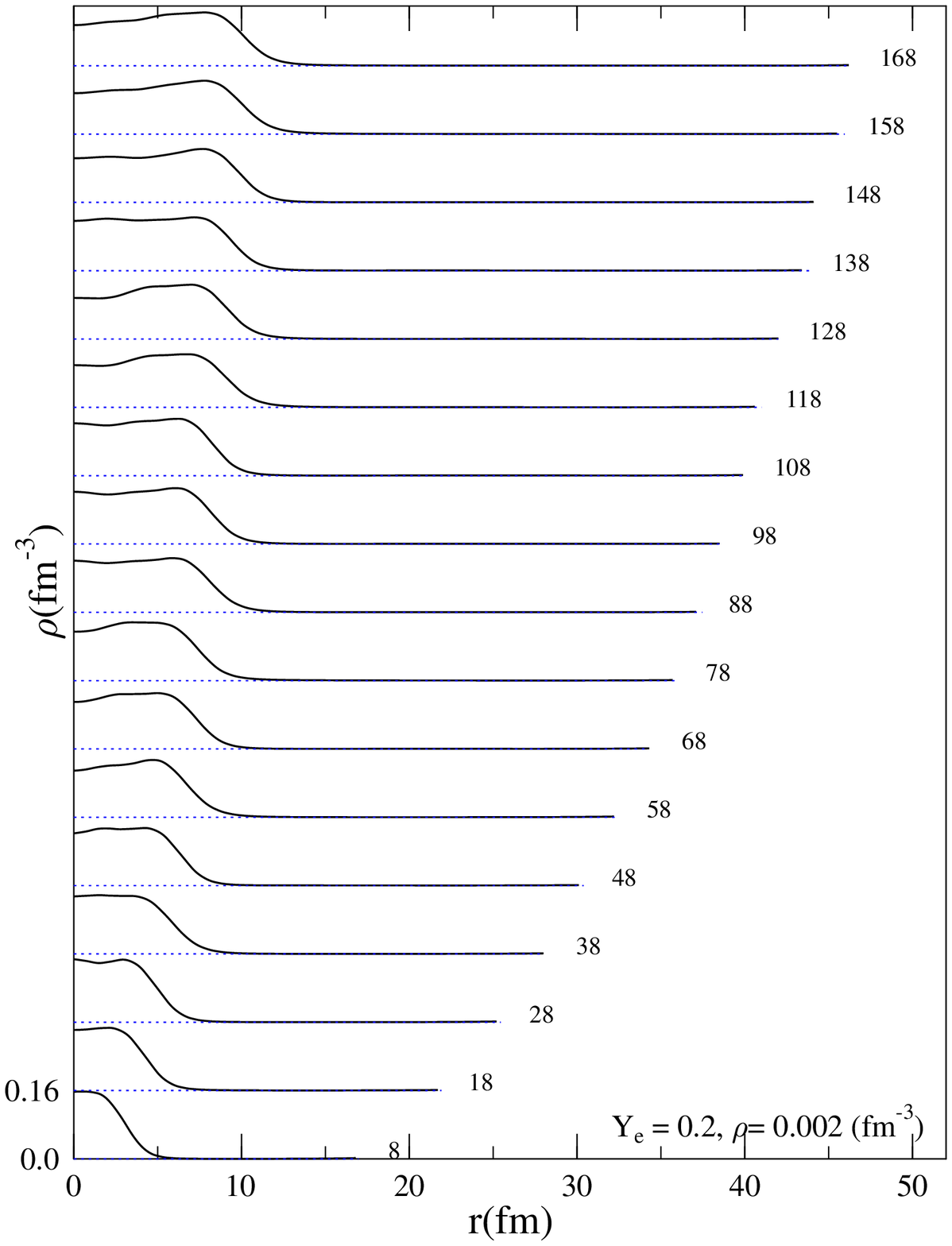}
    \caption{%
        Baryon densities for a sequence of nuclei with increasing
        $\gls{num_byn}$ and fixed charge-to-baryon ratio of $\gls{rat_ptb}
        = 0.2$ at average baryon density of $\gls{adens_byn} = \SI{.002}{\per
        \f \cubed}$. The distributions are extended up to the corresponding
        $\gls{rad_wsc}$. The corresponding proton numbers are indicated at the
        curves.}
    \label{fig:const_y_1}
\end{figure*}
\begin{figure*}[t]
    \centering
    \includegraphics[width=.8\textwidth]
    {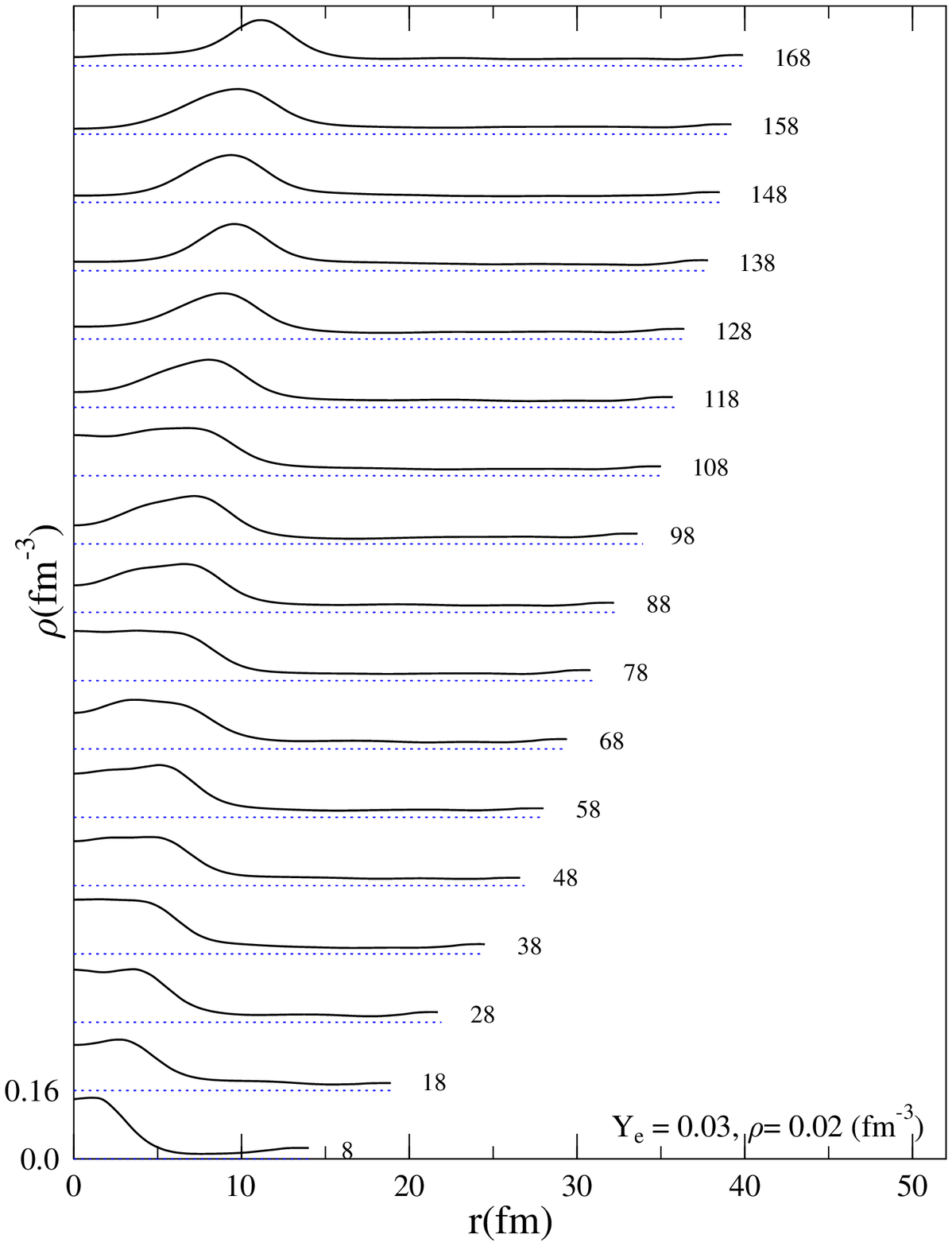}
    \caption{%
        The same as in \fref{const_y_1} but for $\gls{rat_ptb} = 0.03$ and
        average baryon density of $\gls{adens_byn} = \SI{.02}{\per \f
        \cubed}$.}
    \label{fig:const_y_2}
\end{figure*}
In this subsection we present calculations for sequences of nuclei with
constant proton-to-baryon ratio
\begin{equation}
    \gls{rat_ptb} = \frac{\gls{num_prn}}{\gls{num_byn}}
\end{equation}
which corresponds also to the fixed electron fraction in the charge-neutral
Wigner-Seitz cell. Such constraint is often used to specify physical conditions
in stellar matter.
\par
In \fref{const_y_1} we show the baryon density profiles for $\gls{rat_ptb}
= 0.2$ and an average baryon density in the cell of $\SI{.002}{\per \f
\cubed}$, which is about 75 times lower than the normal nuclear density. For
$\gls{num_prn} = 8$ (the bottom panel in the figure), the corresponding neutron
number is $\gls{num_ntn} = 32$. The top panel corresponds to proton number
$\gls{num_prn} = 168$ and neutron number $\gls{num_byn} = 672$. For each panel,
the dashed line indicates zero density. One can see that the profiles evolve
slowly from the Woods-Saxon type of distributions at $\gls{num_prn} \lesssim
50$ to more hollow shapes at larger $\gls{num_prn}$.
\par
This trend becomes even more pronounced for higher baryon density
$\gls{dens_byn} = \SI{.02}{\per \f \cubed}$ and smaller $\gls{rat_ptb} = 0.03$,
as shown in \fref{const_y_2}. In this case the hollow shape gradually evolves
to the extreme configuration where nucleons are located on a spherical shell
with a width of about \SI{5}{\f}. As one can see, the central density in this
case is not much different from the density near the WS cell boundary. Again,
the dashed line indicates the zero density level. In \fref{const_y_2}, the
proton number $\gls{num_prn} = 8$ and neutron number $\gls{num_ntn} = 259$ are
chosen for the bottom panel and proton number $\gls{num_prn} = 168$ and neutron
number $\gls{num_ntn} = 5432$ are fixed in the top one. One should bear in mind
that these nuclei are generally not in $\beta$-equilibrium.
\subsection{Single-particle states}
\subsubsection{$\beta$-stable isotopes}
\begin{figure*}[t]
    \centering
    \includegraphics[width=.8\textwidth]
    {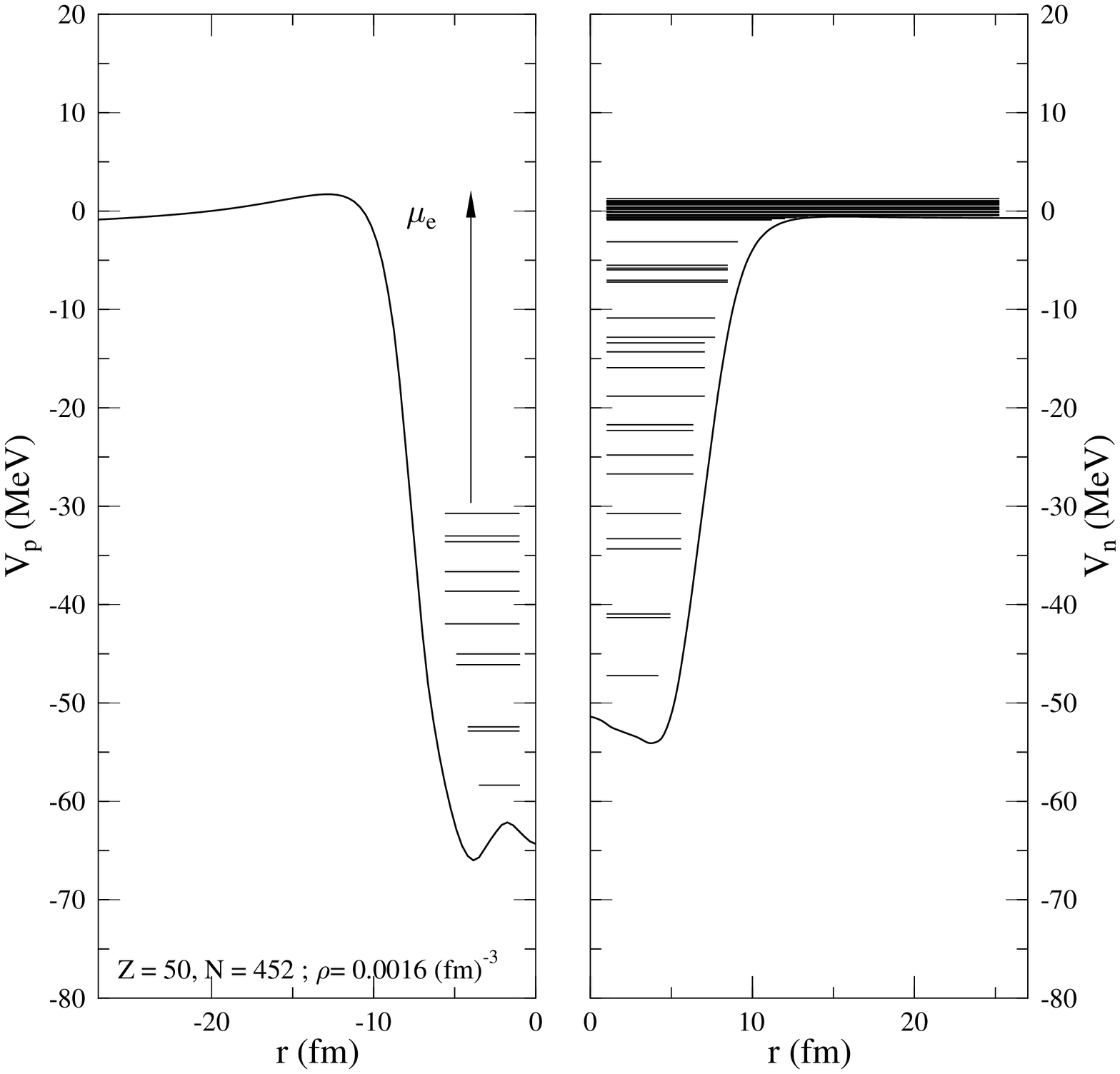}
    \caption{%
        Occupied single-particle states of protons (left) and neutrons (right)
        for the system with $\gls{num_prn} = 50$, $\gls{num_ntn} = 452$ at
        $\gls{adens_byn} = \SI{.0016}{\per \f \cubed}$. The arrow shows the
        electron chemical potential which is equal to the energy difference at
        the upper neutron and proton levels.}
    \label{fig:50_452}
\end{figure*}
\begin{figure*}[t]
    \centering
    \includegraphics[width=.8\textwidth]
    {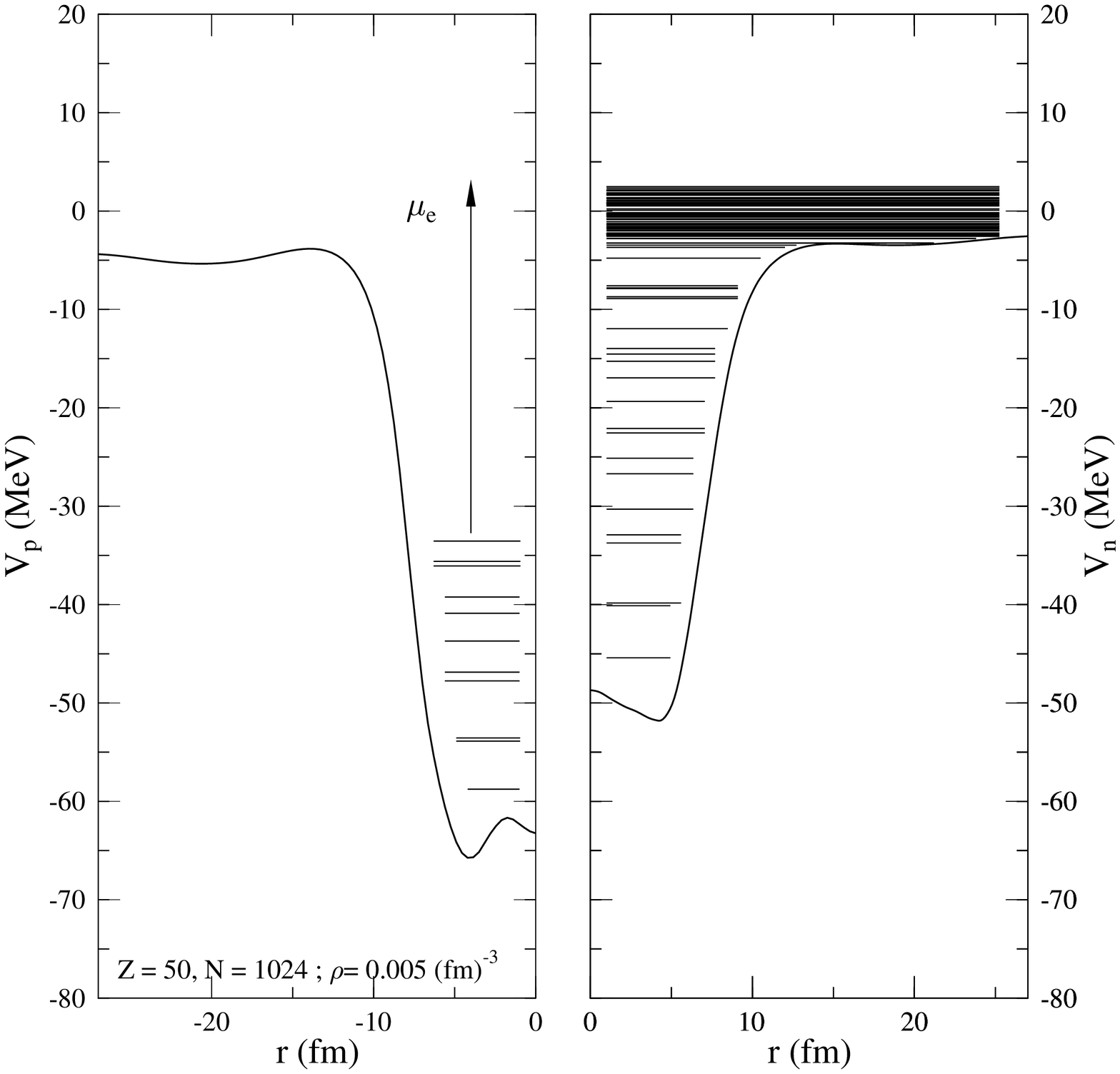}
    \caption{%
        The same is in \fref{50_452} but for the system with $\gls{num_prn}
        = 50, \gls{num_ntn} = 1024$ at $\gls{adens_byn} = \SI{.005}{\per \f
        \cubed}$.}
    \label{fig:50_1024}
\end{figure*}
\begin{figure*}[t]
    \centering
    \includegraphics[width=.8\textwidth]
    {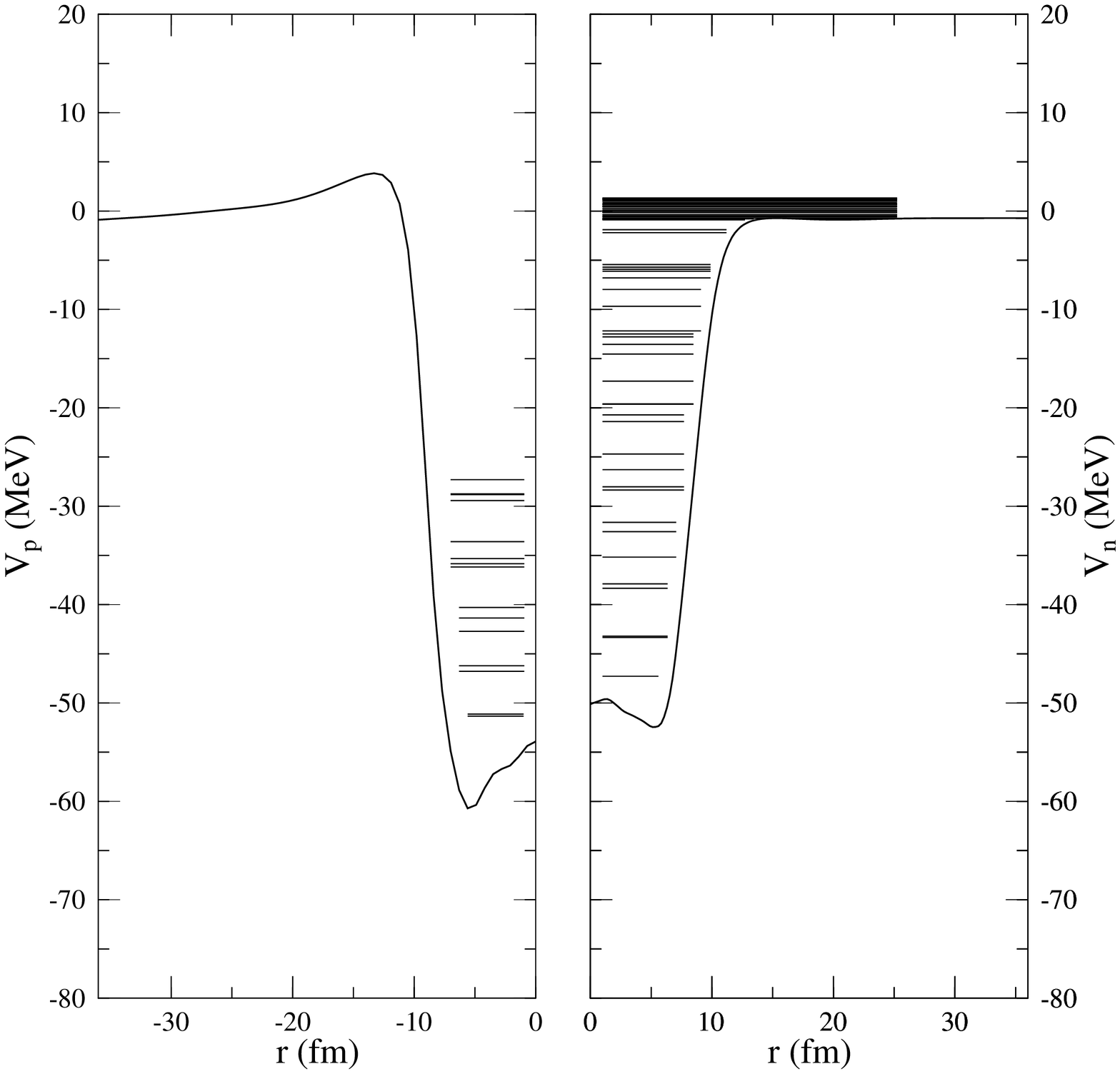}
    \caption{%
        The same is in \fref{50_452} but for the system with $\gls{num_prn}
        = 82, \gls{num_ntn} = 944$ at $\gls{adens_byn} = \SI{.0016}{\per \f
        \cubed}$.}
    \label{fig:82_944}
\end{figure*}
Let us consider now the $\beta$-equilibrium conditions in WS cells which
contain unbound neutrons and an uniform electron background. At fixed average
baryon density $\gls{adens_byn}$, such systems can be specified by the total
charge (isotopes) or baryon number (isobars). Then, the volume of the cell is
obtained self-consistently by the iterative procedure. In \fref{50_452}, the
occupied single particle states for protons (left) and neutrons (right) are
shown within the respective single-particle potential for the system with
$\gls{num_prn}=50, \gls{num_ntn}=452$, which is the $\beta$-stable tin isotope
for the average baryon density $\gls{adens_byn} = \SI{.0016}{\per \f \cubed}$.
The arrow shows the electron chemical potential, which is added to the proton
chemical potentialto reach the upper occupied level of neutrons, see
\eref{beta_stab}.
\par
The occupied proton states are well bound in a potential well that resembles
the proton potential for ordinary nuclei, with the exception that it does not
go to zero as $r$ increases but reaches a constant (negative) value. This is
due to the constant neutron and electron backgrounds which generate an
attractive potential. The lowest neutron levels are well bound in the potential
well. At the energy level of about \SI{-1}{\me}, the neutron potential becomes
flat up to the cell boundary. At this point, a drastic increase of the level
density is observed.
\par
While for ordinary nuclei the proton potential is less deep compared to the
neutron potential, we encounter an opposite situation here. Since we require
charge neutrality of the Wigner-Seitz cell, the cell is filled with electrons
which interact attractively with the protons, and do not directly interact with
the neutrons. There is an indirect effect due to the self-consistency of
nuclear interactions, but it is rather small. Hence, due to the attractive
electron background, the repulsion of the protons is reduced compared to the
situation in vacuum.
\par
The $\beta$-stable tin isotope that corresponds to the baryon density
$\gls{adens_byn} = \SI{.005}{\per \f \cubed}$ is shown in \fref{50_1024}. Since
the baryon density is larger than in the previous case, the radius and
correspondingly the volume of the WS cell is smaller. Hence the electron
chemical potential is larger, and in order to reach $\beta$-equilibirum, more
neutrons should be added to the cell. Consequently, neutrons in this nucleus
occupy higher-lying states.
\par
Due to the selfconsistency of the calculation, especially due to the
interaction of protons with neutrons and electrons, the proton single-particle
states in the WS cell with $\gls{num_prn} = 50$, $\gls{num_ntn} = 1024$ have
shifted down, but the neutron states have shifted up. The general picture is
not very different for a heavier system with $\gls{num_prn} = 82$,
$\gls{num_ntn} = 944$ shown in \fref{82_944}.
\subsubsection{$\beta$-stable isobars}
\begin{figure*}[t]
    \centering
    \includegraphics[width=.8\textwidth]
    {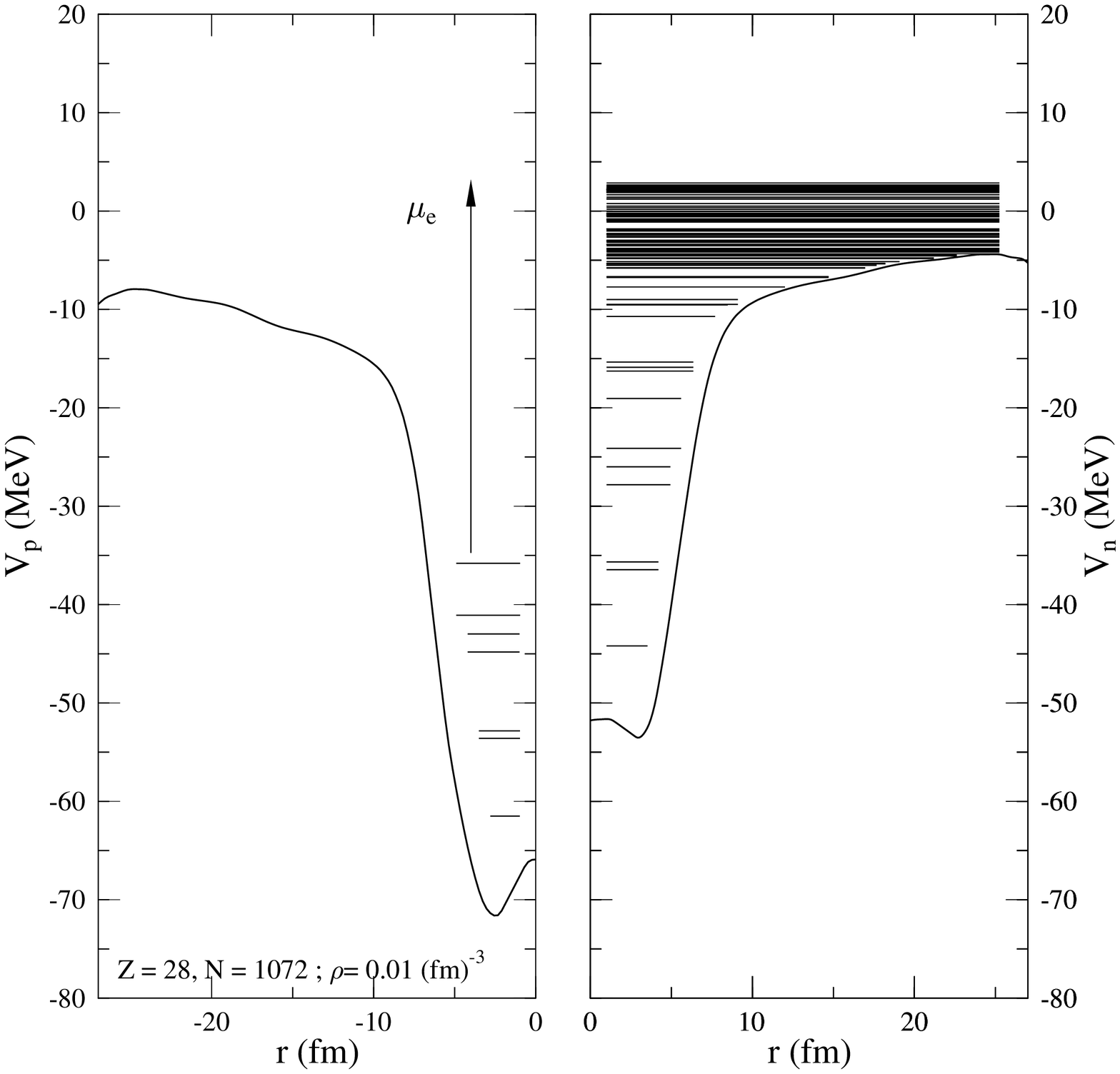}
    \caption{%
        The same is in \fref{50_452} but for the system with $\gls{num_prn}
        = 28, \gls{num_ntn} = 1072$ and $\gls{adens_byn} = \SI{.01}{\per \f
        \cubed}$. This nucleus is the $\beta$-stable isobar of $\gls{num_byn}
        = 1100$ for the specified density.} 
    \label{fig:28_1072}
\end{figure*}
\begin{figure*}[t]
    \centering
    \includegraphics[width=.8\textwidth]
    {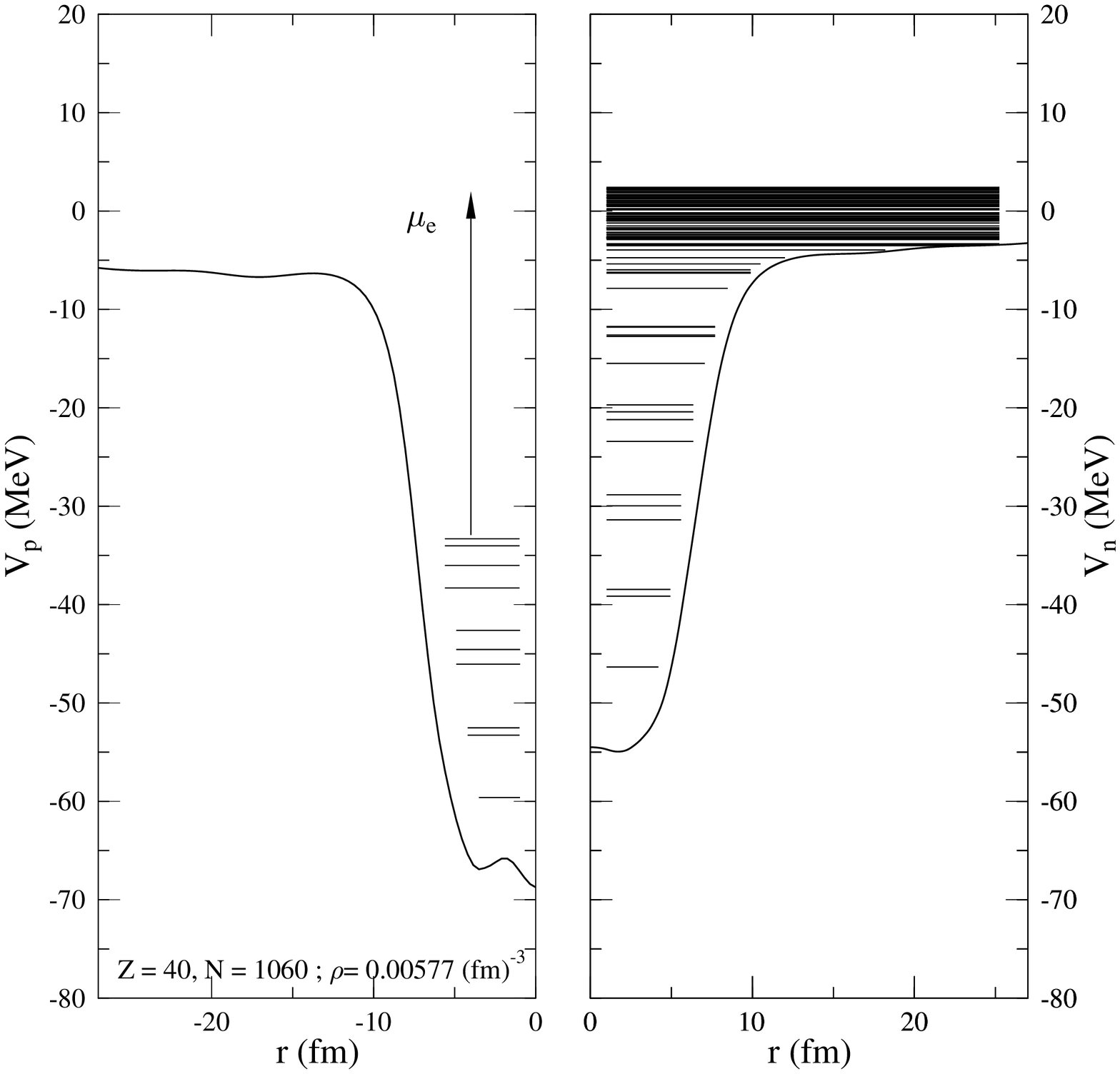}
    \caption{%
        The same is in \fref{50_452} but for the system with
        $\gls{num_prn}=40$, $\gls{num_ntn}=1060$ and $\gls{adens_byn}
        = \SI{.00577}{\per \f \cubed}$. This nucleus is the $\beta$-stable
        isobar of $\gls{num_byn} = 1100$ for the specified density.}
    \label{fig:40_1060}
\end{figure*}
It is interesting to see how the nuclear structure changes with increasing
baryon density for isobars, i.e. at fixed $\gls{num_byn}$ and different
$\gls{num_prn}$. Two isobars of $\gls{num_byn} = 1100$ are presented in
Figs.~\ref{fig:28_1072} and \ref{fig:40_1060} for baryon densities
$\gls{adens_byn} = \SI{.01}{\per \f \cubed}$ and $\gls{adens_byn}
= \SI{.00577}{\per \f \cubed}$, respectively. Since the electron chemical
potential rises with the density, more protons convert to neutrons at higher
baryon densities to maintain the $\beta$-equilibrium. As one can see, indeed at
the larger baryon density shown in \fref{28_1072}, the system has larger
$\gls{num_ntn}$ and smaller $\gls{num_prn}$ compared to the one shown in
\fref{40_1060}.
\section{Nuclear structure calculations with realistic electron background}
\subsection{Electron density distributions}
\label{ss:tf}
In this section, we extend the model by replacing the constant electron density
with a realistic, nonuniform density distribution obtained self-consistently by
using the relativistic generalization of the Thomas-Fermi method. In the
quasiclassical approximation one can write the energy of an electron of
momentum $\vec{k}$ at point $r$ as
\begin{equation}
    E (\vec{k},r) = \sqrt{ \vec{k}^2 + \gls{mass_eln}^2 } - e \phi (r)
    \text{.}
\label{eq:tf:fermi-energy}
\end{equation}
All states with $E(\vec{k},r) \leq \gls{cpot_eln}$ are occupied, where
$\gls{cpot_eln}$ is the electron chemical potential independent of $r$. The
local Fermi momentum $\gls{f_mom}( r )$ is determined by the condition
\begin{equation}
    \gls{cpot_eln} = \sqrt{ \gls{f_mom}^2( r ) + \gls{mass_eln}^2 } - e \phi(
    r )
    \text{,}
\end{equation}
hence
\begin{equation}
    \gls{f_mom} (r) = \sqrt{ \left( \gls{cpot_eln} + e \phi (r) \right)^2
    - \gls{mass_eln}^2 }
    \text{.}
\label{eq:tf:fermi-momentum}
\end{equation}
Summing all the electron states from $k = 0$ up to the local Fermi momentum
$\gls{f_mom}(r)$ we get the local electron density
\begin{equation}
    \gls{dens_eln}( r )
    = \nu_e \int^{ \gls{f_mom} ( r ) }_0 \frac{ \diff^3 k }{(2 \pi )^3}
    = \frac{ \gls{f_mom}^3 ( r )}{ 3 \pi^2 }
    = \frac{ 1 }{ 3 \pi^2 } \left[ \left( \gls{cpot_eln} + e \phi(
    r ) \right)^2 - \gls{mass_eln}^2 \right]^{\frac{ 3 }{ 2 }}
    \text{,}
\label{eq:rho}
\end{equation}
where $\nu_e = 2$ is the spin degeneracy factor. The requirement of
electroneutrality of the cell implies that the total electron number is equal
to the total proton number $Z$ in the cell volume
$\gls{vol_wsc}$,
\begin{equation}
    \int_{\gls{vol_wsc}} \gls{dens_eln}( r ) \diff^3 r = \gls{num_prn}
    \text{,}
    \label{eq:tf:charge-conservation}
\end{equation}
which determines the electron chemical potential $\gls{cpot_eln}$.
\par
The proton density is obtained from the RMF calculation, so the electrostatic
potential can be determined self-consistently from the Poisson equation
\begin{equation}
    \laplace \phi = - \gls{dens_tc}( r )
    \text{,}
    \label{eq:poisson}
\end{equation}
where $\gls{dens_tc}$ is the total charge density defined in \eref{dens_tc}
which itself depends on $\phi( r )$ via eq.~\eqref{eq:rho}. This equation is
solved with the boundary conditions for $\phi( r )$ requiring regularity at the
origin and vanishing derivative at the cell boundary. The nonuniform electron
density distribution calculated in this way determines the electrostatic
potential for the next iteration of the RMF calculation. All particle
densities, including electrons, are obtained self-consistently. In order to use
the same notation as in the previous sections, we define $k_F$ (without
argument $r$) to characterize the average electron density in the WS cell, 
\begin{equation} \label{eq:aed} 
    \gls{adens_eln}
    = \frac{\gls{num_prn}}{\gls{vol_wsc}}
    = \frac{\gls{f_mom}^3}{3 \pi^2}
    \text{.}
\end{equation}
\subsection{Numerical implementation of self-sonsistent approach}
Now we are going to study the properties of nuclei embedded in the
realistic electron background, in particular, their binding energies as
well as neutron and proton drip lines. To fulfill this task one should
evaluate a broad ensemble of particle-stable nuclei, i. e. the whole
nuclear chart. Certainly, this task presents a serious computational
challenge. Below we limit ourselves to the nuclei with even proton numbers
from 8 to 240 and neutron numbers from 6 to 360. Then the total number of WZ
cells which should be coinsidered amounts to about 60000. Each individual cell
calculation takes a CPU time between minutes and hours, depending on the
electron Fermi momentum $\gls{f_mom}$ and the nucleon number. For these
extensive computations we have used the CPU cluster at LOEWE Center for
Super-Computing (CSC) at Goethe University. We carefully selected appropriate
configurations for the SLURM workload manager used at this cluster and wrote
special scripts to assign the amount of cells to allow parallel computation.
\par
Auxiliary scripts created the input files for each WS cell system calculation
and assigned the calculations to the corresponding computer nodes. The
iteration scheme for each system started with the calculation of the baryon
densities within the RMF description, where the solution of the Dirac equation
gave the single-particle wave functions. The many-particle wave function was
then obtained by calculating the Slater determinant. We used cubic B-Splines to
interpolate between the numeric grid of the nucleons and the electrons. The
Thomas-Fermi routine calculated the self-consistent electrons by iterating
between the solution of the one-dimensional Poisson \eref{poisson} and the
electron density distribution given by \eref{rho}. The numerical scheme of the
Thomas-Fermi routine is described in \cite{Ebel2015}.
\subsection{Modification of nuclear chart under stellar conditions}
\begin{figure*}
    \centering
    \includegraphics[width=.8\textwidth]
        {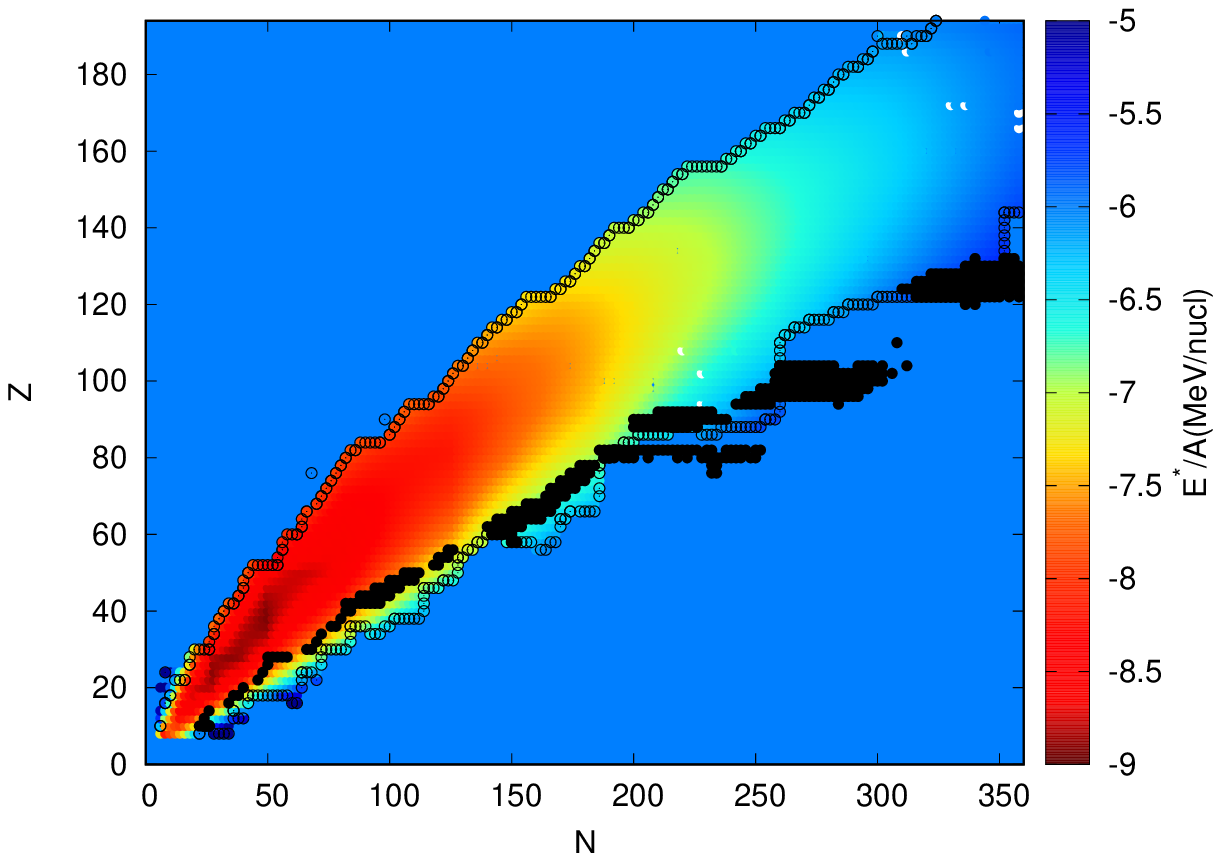}
    \caption{%
        The energy per baryon including electron kinetic energy $E^{\ast}$ for
        $k_F = \SI{.1}{\per \f}$. Shown are the proton dripline (black
        circles), the beta stability line (black dots) and the neutron dripline
        (black circles).}
    \label{fig:en} 
\end{figure*}
\begin{figure*}
    \centering
    \includegraphics[width=.8\textwidth]
        {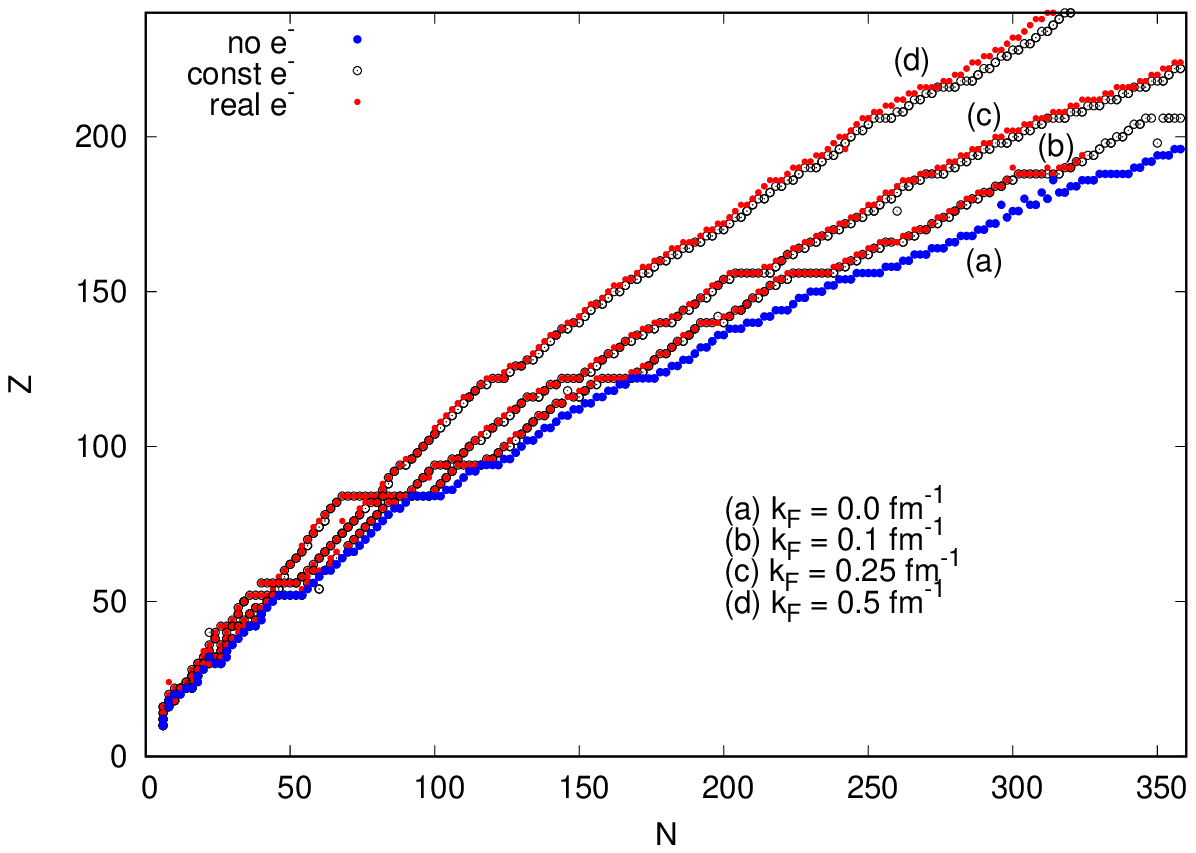}
    \caption{%
        The proton driplines for various electron Fermi momenta with no
        electrons (blue dots), constant (black circles) and realistic (red
        dots) electron distribution.}
    \label{fig:dl} 
\end{figure*}
\begin{figure}
\begin{center}
\centering
    \includegraphics[width=.8\textwidth]
        {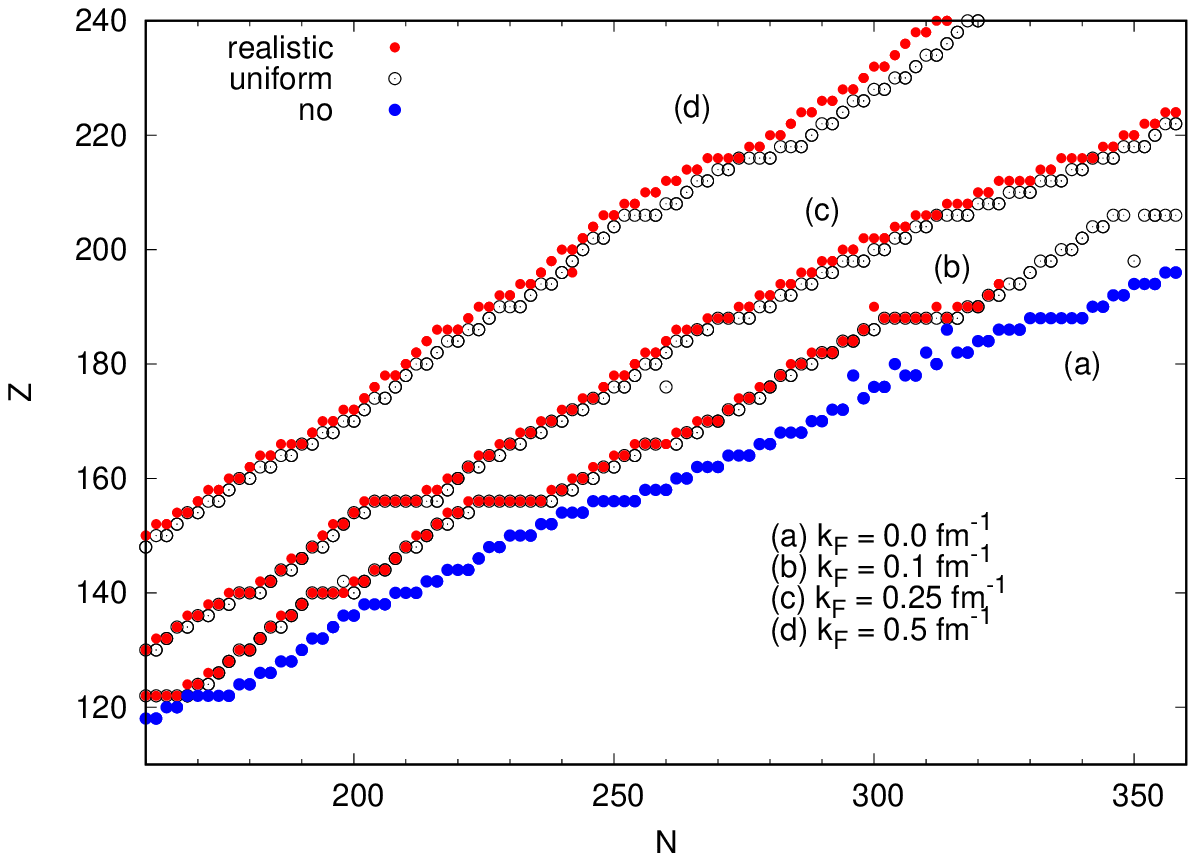}
\caption{%
    The proton driplines of the heaviest nuclei for various electron fermi
    momenta with no electrons (blue dots), constant (black circles) and
    realistic (red dots) electron distribution.}
\label{fig:dlz} 
\end{center}
\end{figure}
In this subsection, we present results of our nuclear structure calculations
for characteristic conditions in outer crusts of neutron stars. They can be
conveniently specified by the average electron density $\gls{adens_eln}$ or the
Fermi momentum $\gls{f_mom}$ as defined in \eref{aed}. We believe that rather
heavy nuclei can exist in the crusts of neutron stars or supernova interiors at
subnuclear densities and low temperatures. Numerous calculations within the
statistical approach predict broad distributions of nuclei up to mass numbers
400-500, see e.g. recent paper \cite{Furusawa2017} and references therein. As
was demonstrated in our previous work, the spontaneous fission of such nuclei
is suppressed because of the screening effect of electrons.
\par
Our particular interest was to compare results obtained in \cite{Burvenich2007}
within a simplified model of uniform electron background with more realistic
calculations presented in this paper. To better illustrate the difference, we
calculate the energy per nucleon defined as
\begin{equation}
    \gls{en_wsc} = \gls{en_nucl} + \gls{en_kin} - \gls{en_kin_un}
    \text{\,,}
\end{equation}
where $\gls{en_nucl}$ is the total energy of the nucleus (with respect to the
rest mass mass of nucleons), and $\gls{en_kin} - \gls{en_kin_un}$ is the change
of the electron kinetic energy between realistic (r) and constant (c) density
distributions in each cell. This change is significant for heavy nuclei, as we
have shown in \cite{Ebel2015}.
\par
In \fref{en}, the results are presented for the electron Fermi momentum of
$\gls{f_mom} = \SI{0.1}{\per \f}$. One can see that the energy gain per nucleon
ranges from \SI{-5}{\me}  for light to \SI{-9}{\me} for heavy nuclei. Beyond
the proton dripline, in addition to nuclei and electrons, the cells will
contain also free neutrons. It is interesting to note, that the valley of the
most bound nuclei in general does not coincide with the beta stability line,
defined by the condition on the electron chemical potential $\gls{cpot_eln}
= \gls{cpot_ntn} - \gls{cpot_prn}$.
\par
In \fref{dl} we present the proton drip lines for a wider range of electron
densities from $\gls{f_mom} = \SI{0}{\per \f}$ to $\SI{0.5}{\per \f}$. They are
compared with the previous calculations \cite{Burvenich2007} for constant
electron densities. The drip lines are defined in terms of the separation
energies $\gls{en_sep_prn} ( \gls{num_ntn}, \gls{num_prn} ) = \gls{en_bind}
( \gls{num_ntn}, \gls{num_prn} ) - \gls{en_bind} ( \gls{num_ntn}, \gls{num_prn}
- 1 )$ which is approximated by a half of the two-proton drip line,
$\gls{en_sep_prn} (\gls{num_ntn}, \gls{num_prn}) \approx (\gls{en_bind}
(\gls{num_ntn}, \gls{num_prn}) - \gls{en_bind} (\gls{num_ntn}, \gls{num_prn}
- 2)) / 2$. By definition, the flip of the sign marks the position of the
dripline.
\par
As expected, the effect of realistic electron distributions is strongest for
most heavy nuclei and highest Fermi momentum of electrons. For instance, for
$\gls{f_mom} = \SI{0.5}{\per \f}$, we have found for $\gls{num_ntn} = 312$
a shift of the proton dripline from $\gls{num_prn} = 234$ to $\gls{num_prn}
= 240$. At smaller $\gls{f_mom}$, the shift varies from 1 to 2 units of charge.
Still, even for $\gls{f_mom} = \SI{0.1}{\per \f}$, the shift can reach a few
charge units, e.g. for $\gls{num_ntn} = 324$ from $\gls{num_prn} = 192$ to
$\gls{num_prn} = 194$, as shown in \fref{dlz}. Similar shifts are also
predicted for the charge and mass numbers of most bound nuclei.
\par
\begin{figure*}
\begin{center}
\includegraphics[width=.8\textwidth]
    {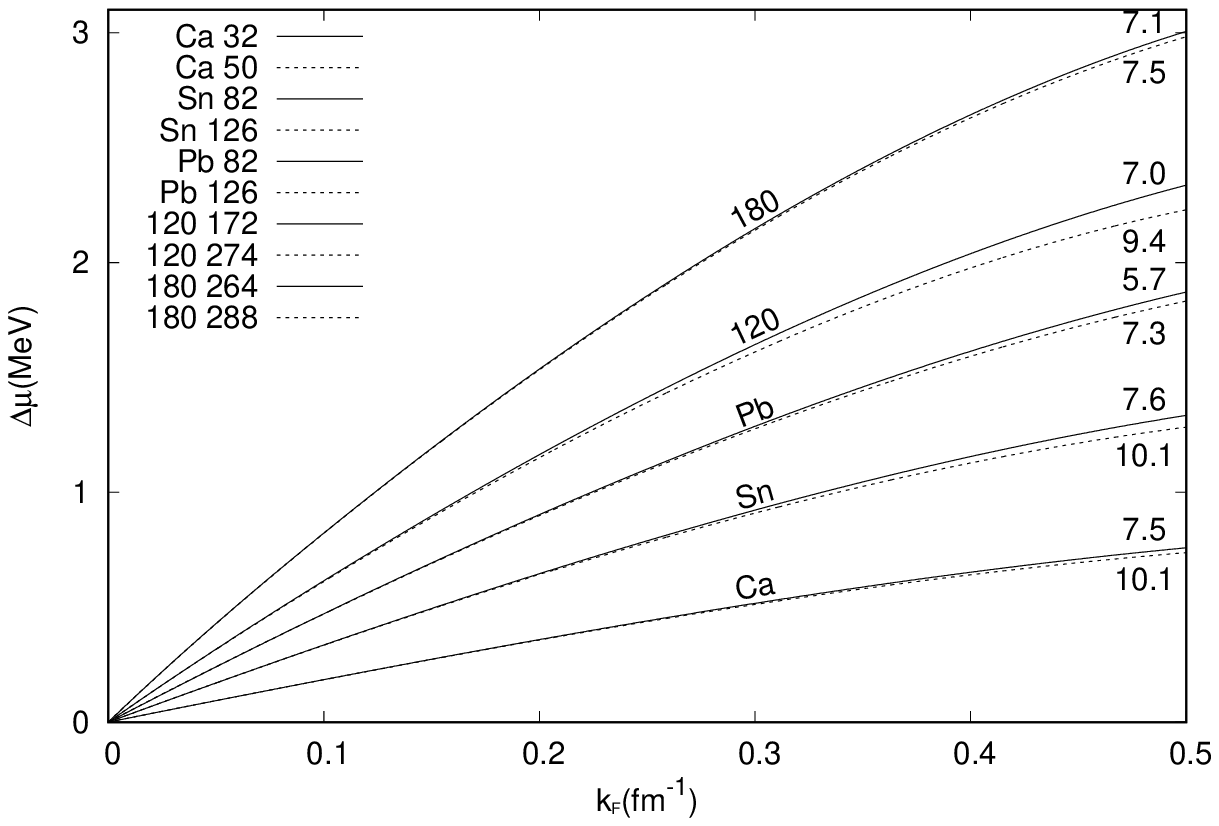}
\caption{%
    The electron chemical potential difference $\Delta \mu$ as a function of
    Fermi momentum $k_F$ for different elements and isotopes for $k_F
    = \SI{.5}{\per \f}$. The corresponding cell filling fractions $\langle
    \rho_{B} \rangle / \rho_0$ in percent are shown on the right at each
    curve.}
\label{fig:mu}
\end{center}
\end{figure*}

As demonstrated in ref.~\cite{Ebel2015}, using realistic electron distributions
is very important for calculating the electron chemical potential
$\gls{cpot_eln}$. The correction depends on the average baryon density
$\gls{adens_byn}$ as well as the electron-to-baryon fraction $\gls{eln_frac}$.
\Figref{mu} shows the shift of the electron chemical potential
$\gls{cpot_eln_diff} = \gls{cpot_eln_un} - \gls{cpot_eln}^\mathrm{r}$ for five
different proton numbers (\gls{num_prn}$ = $ \numlist{20;50;82;120;180}) with
two isotopes each, calculated for varying electron Fermi momenta up to $k_F
= \SI{.5}{\per \f}$.
\par
Following ref.~\cite{Ebel2015}, we define the cell filling fraction as
\begin{equation} \label{eq:cff} 
    \gls{ffrac_wsc}
    = \frac{\gls{adens_byn}}{\gls{dens_nnucl}}
    = \frac{\gls{dens_eln}}{\gls{eln_frac} \gls{dens_nnucl}}
    = \frac{\gls{f_mom}^3}{3 \pi^2 \gls{eln_frac} \gls{dens_nnucl}}
    \text{,}
\end{equation}
where $\gls{adens_byn}$ is the mean baryon density und $\rho_0$ is the normal
nuclear density as in \cite{Ravenhall1983}. In order to compare our results
with our previous calculations in ref.~\cite{Ebel2015}, we show the values of
the cell filling fraction $\gls{ffrac_wsc}$ (in percent) for each isotope curve
on the right side of \fref{mu}. The numbers are given by the right side of
\eref{cff} for $k_F = \SI{.5}{\per \f}$ and for $\rho_0 = \SI{.147}{\per \f}$
(as used in refs.~\cite{Ebel2015} and~\cite{Ravenhall1983}). Obviously, the
electron-to-baryon ratio $\gls{eln_frac}$ for individual cells depends on the
isotope composition. Indicated in \fref{mu} are the cell filling fractions
below $\SI{10}{\percent}$. In this case, considering the nucleus in the shape
of a droplet is still justified, see ref.~\cite{Ebel2015}. At $\gls{ffrac_wsc}
> 0.2$, the nuclear matter distribution is described by pasta phases
\cite{Lattimer1985}.
\section{Conclusions}
In the first part of this paper, we have performed nuclear structure
calculations in Wigner-Seitz cells with uniform electron background. The
calculations were carried out using the relativistic mean-field model with the
NL3 parametrization. Physically motivated boundary conditions were introduced
for the neutron single-particle states to describe unbound neutrons. We have
studied the structural evolution of shapes along isotopic chains until very
high neutron-rich systems and for fixed charge-to-baryon ratios. For superheavy
isotopes, the generation of hollow shapes with a central deep in the proton
distribution was found. We have also studied the single-particle states and
mean-field potentials for different nuclei embedded in the neutron end electron
background.
\par
In the second part of the paper, we have replaced the constant electron
background with realistic nonuniform distributions, found from the Poisson
equation. The self-consistent calculations of nuclear and electric potentials
were done within an iterative scheme, using the RMF description for the
nucleons and the Thomas-Fermi approximation for the electrons. The differences
between the electron chemical potentials for uniformly and nonuniformly
distributed electrons were calculated for different atomic numbers and
isotopes.
\par
The whole nuclear chart was calculated for even numbers of protons between
$\gls{num_prn} =$ 8 and 240 and neutrons between $\gls{num_ntn} = 6$ to
$\gls{num_ntn} = 360$, both for realistic and uniform electron distributions.
We have systematically compared the energies per baryon for these two cases and
found the beta stability line as well as the neutron- and proton driplines.
Additionally, we calculated the proton driplines for both cases and for the
electron Fermi momentum of $\gls{f_mom} = \SI{0.25}{\per \f}$ and $\gls{f_mom}
= \SI{0.5}{\per \f}$, as well as for vacuum. At high electron densities, we
have found a shift of several charge units in the proton dripline, caused by
the realistic (non-uniform) electron density distributions. Despite the fact
that all nuclear structure calculations in this paper were performed with RMF
parametrization NL3, we believe that the main predictions will be similar for
other popular versions of the RMF model.
\section{Acknowledgements}
This work is devoted to the memory of Prof. Walter Greiner, who was a great
scientist, great teacher and great man. We started this work about 5 years ago
after many fruitful discussions with him. We express our sincere gratitude for
his continuous interest to this work and valuable advices. We also thank S.
Schramm and U. Heinzmann for fruitful discussions. This work was supported in
part by the Helmholz International Center for FAIR research.
\bibliography{library.tex}
\end{document}